\documentclass[prd,showpacs,twocolumn]{revtex4-1}

\usepackage{amsmath}
\usepackage{amsfonts}
\usepackage{graphicx}
\usepackage[table]{xcolor}
\newcommand{\gc}{\boldmath\cellcolor{lightgray}}

\renewcommand{\bar}{\overline}

\newcommand{\CC}{{\mathbb C}}
\newcommand{\HH}{{\mathbb H}}
\newcommand{\KK}{{\mathbb K}}
\newcommand{\OO}{{\mathbb O}}
\newcommand{\RR}{{\mathbb R}}

\newcommand{\UU}{1}

\renewcommand{\sl}{\mathfrak{sl}}
\newcommand{\so}{\mathfrak{so}}
\newcommand{\su}{\mathfrak{su}}
\renewcommand{\sp}{\mathfrak{sp}}
\newcommand{\spin}{\mathfrak{spin}}
\newcommand{\uu}{\mathfrak{u}}
\renewcommand{\aa}{\mathfrak{a}}
\newcommand{\dd}{\mathfrak{d}}
\newcommand{\ee}{\mathfrak{e}}
\newcommand{\ff}{\mathfrak{f}}
\renewcommand{\gg}{\mathfrak{g}}
\newcommand{\hh}{\mathfrak{h}}

\newcommand{\suL}{\su2_L}
\newcommand{\suR}{\su2_R}
\newcommand{\suc}{\su3_c}

\newcommand{\Spin}{\mathrm{Spin}}
\newcommand{\SO}{\mathrm{SO}}
\newcommand{\SU}{\mathrm{SU}}

\newcommand{\eA}{A}	
\newcommand{\eG}{G}	
\newcommand{\eS}{S}	
\newcommand{\eGS}{GS}	
\newcommand{\eD}{D}	
\newcommand{\eX}{X}	
\newcommand{\eY}{Y}	
\newcommand{\eZ}{Z}	

\newcommand{\up}{\uparrow}
\newcommand{\dn}{\downarrow}
\newcommand{\Ru}{R\up}
\newcommand{\Rd}{R\dn}
\newcommand{\Lu}{L\up}
\newcommand{\Ld}{L\dn}

\renewcommand{\Im}{{\mathrm{Im}}}
\renewcommand{\Re}{{\mathrm{Re}}}
\newcommand{\tr}{\mathrm{tr}}
\newcommand{\Cl}{\mathrm{C}\ell}
\newcommand{\bpsi}{{\boldsymbol\psi}}

\newcommand{\mb}[1]{\mathbf{#1}}
\newcommand{\HY}{\eta}		
\newcommand{\HYp}{\HY_{\mathrm{perp}}}
\newcommand{\Sp}{\mathbf{S}}	
\newcommand{\eW}{R}

\newcommand{\vp}{\varphi}
\newcommand{\psid}{\vp(\psi)}
\newcommand{\psib}{a(\psi)}

\usepackage{xcolor}
  \definecolor{bluegray}{rgb}{0.4,0.6,0.8}
  \definecolor{forest}{rgb}{0,0.5,0}
  \definecolor{violet}{rgb}{0.33,0,0.67}

\newcommand\Later[1]{}         

\newcommand{\hide}[1]{\relax}
\newcommand\OldLater[1]{}	

\begin{document}

\title{\boldmath Octions: An $E_8$ description of the Standard Model}

\author{Corinne A. Manogue}
\email{corinne@physics.oregonstate.edu}
\affiliation{%
  Department of Physics, Oregon State University, Corvallis, OR  97331, USA}

\author{Tevian Dray}
\email{tevian@math.oregonstate.edu}
\affiliation{%
  Department of Mathematics, Oregon State University, Corvallis, OR  97331,USA}

\author{Robert A. Wilson}
\email{r.a.wilson@qmul.ac.uk}
\affiliation{%
  School of Mathematical Sciences, Queen Mary,
  University of London, London E1 4NS, UK}

\begin{abstract}
We interpret the elements of the exceptional Lie algebra $\ee_{8(-24)}$ as
objects in the Standard Model, including lepton and quark spinors with the
usual properties, the Standard Model Lie algebra
$\su(3)\oplus\su(2)\oplus\uu(1)$, and the Lorentz Lie algebra $\so(3,1)$.  Our
construction relies on identifying a complex structure on spinors and then
working in the enveloping algebra.  The resulting model naturally contains
GUTs based on $\SO(10)$ (Georgi--Glashow), $\SU(5)$ (Georgi), and
$\SU(4)\times\SU(2)\times\SU(2)$ (Pati--Salam).  We then briefly speculate on
the role of the remaining elements of $\ee_8$, and propose a mechanism leading
to exactly three generations of particles.
\end{abstract}


\maketitle

\section{Introduction}
\label{intro}

From the earliest days of quantum mechanics, mathematical physicists have been
searching for a mathematical structure rich enough to describe the physical
world.  Exceptional structures involving the octonions, as the largest of the
normed division algebras, the Albert algebra, as the only exceptional arena
for quantum mechanics, the exceptional Lie algebra $\ee_6$, describing the
symmetries of the Albert algebra, its larger cousins $\ee_7$, and $\ee_8$, and
tensor products of division algebras such as
$\RR\otimes\CC\otimes\HH\otimes\OO$, have all received attention.

Beginning in 1933, Jordan and colleagues~\cite{Jordan,JNW,Albert} showed that
there was just one exceptional, nonassociative algebraic system, now known as
the Albert algebra, in which the postulates of quantum mechanics can be
realized.  However, as described by~\citet[pp.~213--214]{GT} (see
also~\cite{Okubo}), it was not until the 1960s that these structures began to
be used seriously in quantum mechanics.
In the early 1980s, there was an explosion of interest in using octonionic
structures to describe supersymmetry, starting
with~\cite{KugoTownsend83,OliveWest83}, and superstring
theory~\cite{FairlieI,Goddard,FairlieII,Corrigan,SudberyCAM,GSW}; a short
history is given in~\cite{BHsuperI}.
These constructions led us to explore models based on
$E_6$~\cite[e.g.][]{Spin,Dim,York,Denver}.

Meanwhile, in the 1970s, Grand Unified Theories (GUTs) using octonions and
exceptional Lie algebras began to appear, such as the work of G\"ursey and
G\"unaydin~\cite[e.g.][]{Gunaydin,GurseyE6}.  To the best of our knowledge,
\citet{Bars80} were the first to propose a GUT using $\ee_8$.  The beautiful
and extensive work by G\"ursey and collaborators, as summarized in~\cite{GT},
is particularly noteworthy.

More recently, \citet{Lisi} attempted to fit the Standard Model into $\ee_8$,
and \citet{Chester} proposed a model based on decomposing $\ee_8$ (and
provided extensive references to other approaches).  Each of these models
interprets the  elements of $\ee_8$ somewhat differently.
In the 1990s, Dixon (see~\cite{Dixon} and the references cited therein) used
matrices over the tensor product of the four division algebras,
$\RR\otimes\CC\otimes\HH\otimes\OO$, to represent particle states.  More
recently, taking a different tack, Furey and
Hughes~\cite{Furey,Furey18,Furey19,Furey22a,Furey22b} have built a model using
only $\RR\otimes\CC\otimes\HH\otimes\OO$ to represent particle states.  Other
recent work along similar lines includes~\cite{TDV,Krasnov21,Perelman21}.

The Lie group $E_8$ is the largest of the exceptional Lie groups, all of which
are naturally associated with the octonions, the largest of the normed
division algebras.
The Freudenthal--Tits magic square~\cite{Freudenthal,Tits} provides a unified
description of 16 Lie algebras, parameterized by two division algebras (or
their split cousins).  The magic square is remarkable for containing four of
the five exceptional Lie algebras, culminating in $\ee_8$.  Building on
Sudbery's early work~\cite{Sudbery,SudberyChung} using the octonions to
describe Lie algebras over the octonions, Barton and
Sudbery~\cite{SudberyBarton} gave a unified presentation of the magic square
based on Vinberg's symmetric construction~\cite{Vinberg}.  Along the way, they
gave explicit matrix interpretations of the Lie algebras in the first three
rows.  In recent work~\cite{Magic}, we provided a new description of $\ee_8$
in terms of $3\times3$ matrices over (two copies of) the octonions.  We
summarize that description in Appendix~\ref{structure}.

Here, we build up to $\ee_8$ by working our way through the magic square,
offering a physical interpretation at each step that leads naturally to a
description of the physical world that shares many features with the Standard
Model.

Although we work here with the Lie algebra $\ee_8$, earlier efforts by
ourselves~\cite{Lorentz,SO42,2x2,Cubies} and others~\cite{Gunaydin,GT}
to lift the Barton and Sudbery
construction to the group level, combined with the techniques used
in~\cite{Magic}, allow our results to be applied directly at the group level.

\begin{table}
\begin{center}
\begin{tabular}{|c|c|c|c|c|}
\hline
& $\RR$ & $\CC$ & $\HH$ & $\OO$ \\\hline
$\RR'$
 & \gc$\so(3)$ & $\su(3)$ & $\su(3,\HH)$ & $\ff_4$ \\\hline
$\CC'$
 & \gc$\sl(3,\RR)$ & $\sl(3,\CC)$ & \gc$\sl(3,\HH)$ & \gc$\ee_{6(-26)}$ \\\hline
$\HH'$
 & $\sp(6,\RR)$ & $\su(3,3)$ & $\dd_{6(-6)}$ & $\ee_{7(-25)}$\\\hline
$\OO'$	
 & $\ff_{4(4)}$ & $\ee_{6(2)}$ & $\ee_{7(-5)}$ & \gc$\ee_{8(-24)}$\\\hline
\end{tabular}
\end{center}
\caption{The ``half-split'' Freudenthal--Tits magic square of Lie algebras.
The shaded cells show our path through the magic square, as discussed in
Section~\ref{construct}.}
\label{square}
\end{table}

\begin{table}
\begin{center}
\begin{tabular}{|c|c|c|}
\hline
Algebra & Maximal Subalgebra & Centralizer \\\hline
\boldmath$\so(3)$
  & $\so(2)$ & $\gg_2\oplus\gg_{2(2)}$ \\\hline
\boldmath$\sl(3,\RR)$
  & $\so(2,1)\oplus\so(1,1)$ & $\gg_2\oplus\sl(3,\RR)$ \\\hline
\boldmath$\sl(3,\HH)$
  & $\so(5,1)\oplus\so(1,1)\oplus\su(2)$ & $\su(2)\oplus\sl(3,\RR)$ \\\hline
\boldmath$\ee_{6(-26)}$
  & $\so(9,1)\oplus\so(1,1)$ & $\sl(3,\RR)$ \\\hline
\boldmath$\ee_{7(-5)}$
  & $\so(8,4)\oplus\su(2)$ & $\su(2)$ \\\hline
\boldmath$\ee_{8(-24)}$
  & $\so(12,4)$ & --- \\\hline
\end{tabular}
\end{center}
\caption{Selected Lie algebras in the magic square, with their maximal
(proper, reductive) subalgebra in $\so(12,4)$ and their centralizer in
$\ee_8$.}
\label{subalg}
\end{table}

The ``half-split'' version of the magic square~\cite{SudberyBarton} is shown
in Table~\ref{square}, parameterized by a split division algebra $\KK'$ and a
division algebra $\KK$.  We emphasize that we work with \textit{real} Lie
algebras, not complex, unless otherwise noted.  As shown in~\cite{Magic}, the
algebras in the half-split magic square can be interpreted as
$\su(3,\KK'\otimes\KK)$.  Since $\KK'\otimes\KK\subset\OO'\otimes\OO$,
$\ee_{8(-24)}$ contains each of the other algebras in the half-split magic
square as subalgebras.  Some algebras are shown with their common names,
others using the Killing--Cartan classification.  An example of the latter is
the final entry, $\ee_{8(-24)}$, henceforth referred to simply as $\ee_8$.
Here, $8$ denotes the \textit{rank} of the Lie algebra, and $-24$ is the
\textit{signature} of the associated Killing form.
(The rank of a Lie algebra gives the largest number of mutually commuting
basis elements; the signature gives the number of boost-like basis elements
minus the number of rotation-like basis elements, henceforth referred to
simply as ``boosts'' and ``rotations''.)
Physicists beware: there is no factor of~$i$ in our description of Lie
algebras, so infinitesimal rotations are \textit{antisymmetric}, and have
\textit{imaginary} eigenvalues.

We choose a preferred copy of $\dd_8=\so(12,4)\subset\ee_8$ and use labels in
$\OO'\oplus\OO$ to identify the 16 axes, so that pairs of labels identify the
120 basis elements of $\so(12,4)$ as the corresponding rotations and boosts,
regardless of what type of representation (adjoint, vector, spinor) is being
considered.  As discussed further in Section~\ref{construct} and
Appendix~\ref{structure}, the choice of $\so(12,4)\subset\ee_8$ amounts to a
preferred $2+1$ decomposition of the $3\times3$ matrices in
$\ee_8\cong\su(3,\OO'\otimes\OO)$.

As we proceed from upper left to lower right through the magic square, the
algebras become larger, as do their maximal (reductive, proper) subalgebras in
$\so(12,4)$, while their centralizer in $\ee_8$ (the maximal subalgebra with
which they commute) becomes smaller; see Table~\ref{subalg}.  To make a long
story short, the maximal subalgebras contain the symmetry algebras, of which
the remaining elements form spinor representations.

In terms of physical interpretation, our journey starts with a choice of
complex structure, that is, an operator that acts as a complex unit, 
thus allowing us to treat certain subsets of these real algebras as complex
representations.  We then add a $\uu(1)$ that will turn out to be closely
related to hypercharge, then the Lorentz algebra and -- separately -- two
$\su(2)$ subalgebras, then finally $\su(3)$.  At the end of the journey, we
obtain precisely the Lorentz group, an $\SO(10)$ GUT, and just enough spinors
to define one generation of quarks and leptons (and their adjoints).
We provide in
Section~\ref{generations} a possible mechanism for generalizing this
construction to exactly three generations, and in Section~\ref{mediators} a
possible identification of mediators (vector bosons).  All without
leaving~$\ee_8$!

We refer to 
the particles obtained in this unified physical interpretation of (all of) the
elements of $\ee_8$ as \textit{octions}.

Let's begin.

\section{\boldmath Constructing $\ee_8$}
\label{construct}

In this section, we describe our preferred path through the magic square,
discussing which new
physical features are added at each step.  Since our ultimate goal is to
present a unified description in terms of $\ee_8$, each step builds on the
previous ones.  Among other things, this leads us to construct the Cartan
subalgebra of $\ee_8$ as we go.

\subsection{\boldmath $\so(3)$}
\label{so3}

We enter the magic square in the top left corner with the real Lie algebra
$\so(3)$, generated by the three antisymmetric matrices $\eX_1$, $\eY_1$,
$\eZ_1$ (which are defined in Appendix~\ref{structure}).
The Lie algebra $\so(3)$ can be identified with $\su(3,\RR'\otimes\RR)$, using
labels $1\in\RR$, $U\in\RR'$; the subscripts ``$1$'' above are shorthand for
``$1U$.''  The single element~$\eX_1$ generates $\so(2)$, which we choose as
the Cartan subalgebra.  Choosing a Cartan element breaks the symmetry,
yielding a preferred $2\times2$ block structure; as discussed above, we refer
to $\eX_1$ as an infinitesimal rotation in the $1U$-plane.

Explicitly,
\begin{equation}
x\,\eX_1+y\,\eY_1+z\,\eZ_1
  = \begin{pmatrix}0& x& -z\\ -x& 0 &y\\ z& -y& 0\\\end{pmatrix} ,
\end{equation}
with $x,y,z\in\RR$, so that $\eX_1$ can be identified with the $2\times2$
matrix in the upper left of this matrix, and the remaining elements can be
identified with the 2-component column to its right; see
Appendix~\ref{structure}, and especially~(\ref{spinors}).

The elements, $\{\eY_1,\eZ_1\}$ are a basis for a
representation of $\so(2)$.  However, since
\begin{equation}
\bigl[\eX_1,[\eX_1,y\eY_1+z\eZ_1]\bigr] = -(y\eY_1+z\eZ_1)
\end{equation}
we have $\eX_1\circ\eX_1=-1$ in the enveloping algebra on this representation
(as discussed in Appendix~\ref{envelop}), and can therefore interpret $\eX_1$
as a \textit{complex structure} $\iota$ by defining
\begin{equation}
\iota(\psi) = [\eX_1,\psi]
\end{equation}
for any linear combination $\psi$ of $\eY$s and $\eZ$s, thus converting the
$YZ$-subspace of $\so(3)$ into a (complex) spinor representation of
$\so(2)\cong\spin(2)\cong\uu(1)$.  

More generally, a complex structure on any
representation $V$ of a Lie algebra $\hh$ is a linear map $\iota$
from $V$ to itself satisfying
\begin{align}
\iota^2 &= -1 ,
\label{isq}\\
\bigl[X,\iota(\psi)\bigr] &= \iota\bigl([X,\psi]\bigr)
\label{complex}
\end{align}
for any $X\in\hh$ and $\psi\in V$.
In writing the action of $\hh$ on $V$ in~(\ref{complex}) as a commutator, we
are implicitly assuming that both $\hh$ and $V$ are contained in a single,
larger Lie algebra, such as $\ee_8$, an assumption that underlies our entire
program.

\subsection{\boldmath $\sl(3,\RR)$}
\label{sl3R}

We next move down a row in the magic square by adding the label $L\in\CC'$,
thus arriving at $\sl(3,\RR)$, a real form of $\aa_2$ (whose compact real form
is $\su(3)$), consisting of the split, antihermitian cousins of the Gell-Mann
matrices.  Explicitly, $\sl(3,\RR)$ is generated by
$\{\eX_1,\eY_1,\eZ_1,\eX_L,\eY_L,\eZ_L,\eD_L,\eS_L\}$.  As discussed in
Appendix~\ref{structure}, the five basis elements labeled by $L$ are boosts,
and the remaining three are rotations.  The $2\times2$ subalgebra is therefore
$\so(2,1)$, generated by $\{\eX_1,\eX_L,\eD_L\}$, and implementing rotations
and boosts on the labels $\{1,U,L\}$.  The centralizer of $\so(2,1)$ in
$\sl(3,\RR)$ is generated by $\eS_L$, so we can expand the Cartan subalgebra
by including $\eS_L$, which is boost-like.

In this case, the $Y,Z$-subspace splits into two
representations of $\so(2,1)$, namely the eigenspaces of~$\eS_L$, labeled by
$\UU\pm L$.  The eight-dimensional Lie algebra $\sl(3,\RR)$ therefore
decomposes as $\so(2,1)\oplus\langle\eS_L\rangle\oplus(2\times\mb{2})$.
Explicitly, the two spinor $\mb{2}$s are
\begin{equation}
\Sp_\pm = \{y_\pm \eY_{\UU\pm L} + z_\pm \eZ_{\UU\mp L}\}
\label{2s}
\end{equation}
with $y_\pm,z_\pm\in\RR$.
%
(Since $\so(p,q)\cong\spin(p,q)$ as Lie algebras, we use the former notation
even when acting, as here, as infinitesimal $\Spin(p,q)$ transformations.)
Each of $\Sp_\pm$ is a copy of the complex plane, since
\begin{equation}
\iota(\eZ_{\UU\mp L}) = [X_1,\eZ_{\UU\mp L}] = \eY_{\UU\pm L} .
\end{equation}

Furthermore, the Killing form $B$ (see Appendix~\ref{kill}) is degenerate on
each vector space $\Sp_\pm$; in each case, the basis consists of two
orthogonal null elements.  The dual basis of $\Sp_+$ is in $\Sp_-$, and vice
versa, since \hbox{$B(\eY_{\UU\pm L},\eY_{\UU\mp L})=2$}, and similarly for $\eZ$.
Explicitly, the Killing dual of any element in $\Sp_\pm$ is obtained by
reversing the sign of $L$.  This operation $\vp$ of ``$L$ conjugation'' can be
realized on $\Sp_\pm$ in the enveloping algebra of $\so(3,3)$, as discussed in
Appendix~\ref{envelop}; see also Appendix~\ref{kill}.
We interpret Killing duality as an adjoint operation, so that pairs of dual
elements do not represent separate physical degrees of freedom.

\subsection{\boldmath $\sl(3,\HH)$}
\label{sl3H}

Our next stop in the magic square adds the labels $\{i,j,k\}$, thus arriving
at $\sl(3,\HH)\cong\su(3,\CC'\otimes\HH)$, a real form of $\aa_5$ (whose
compact real form is $\su(6)$), with dimension 35.  The $2\times2$ structure
now consists of $\so(5,1)$, acting on the six labels $\{1,i,j,k,U,L\}$ (the
$\eD$s and $\eX$s with labels in $\CC'$ and/or~$\HH$), and there are now 16
spinors (the $\eY$s and $\eZ$s with labels in $\CC'\otimes\HH$).  The
$35-15-16=4$ remaining elements of this basis centralize $\so(5,1)$ in
$\sl(3,\HH)$; these elements include $\eS_L$, of course, but also three new
elements, forming a copy of of $\su(2)$ that we henceforth refer to as $\suR$.

Our previous work~\cite{Magic}, summarized in Appendix~\ref{structure}, shows
that most (but not all) elements of $\ee_8$ admit an explicit matrix form.
Here, since the underlying algebras $\CC'$ and $\HH$ are associative, we can
realize \textit{every} element of $\sl(3,\HH)$ as a matrix.
In particular, we can multiply out the nested double-index $\eD$s, so that,
for example
\begin{equation}
\eD_{i,j} \doteq \begin{pmatrix}k& 0& 0\\ 0& k& 0\\ 0& 0& 0\\\end{pmatrix} ,
\end{equation}
where the dot over the equals sign is to remind us that this expression is not
valid in all of $\ee_8$.  

Similarly, although the elements of $\suR$ must be
represented in $\ee_8$ as (linear combinations of) double-index $\eD$s with
labels in $\HH_\perp=\ell\HH$, within $\sl(3,\HH)$ they take the form
\begin{equation}
\eGS_q \doteq \begin{pmatrix}0& 0& 0\\ 0& 0& 0\\ 0& 0& q\\\end{pmatrix} ,
\end{equation}
with $q\in\Im\HH$.  These two descriptions of $\suR$ can be shown to be
equivalent using triality~\cite{Magic}.
Thus, $\sl(3,\HH)$ can be represented as the $3\times3$ antihermitian matrices
over $\CC'\otimes\HH$, where a nonzero (imaginary) quaternionic trace is
allowed.

The 16 spinor degrees of freedom form two real, 8-dimensional representations,
namely the eigenspaces $\Sp_\pm$ of~$\eS_L$, labeled by $\UU\pm L$, where we
have reused the names introduced in Section~\ref{sl3R}.
Since $L^2=1$,
\begin{equation}
L(\UU\pm L) = \pm (\UU\pm L)  ,
\end{equation}
that is, $\UU\pm L$ are eigenvectors in $\CC'$ of multiplication by~$L$.
Since $\eS_L$ multiplies spinors by (some multiple of) $L$, these eigenspaces
have the matrix form
\begin{equation}
\psi_\pm = \begin{pmatrix}-\bar{q}\\ p\\ \end{pmatrix} (\UU\pm L) ,
\label{line}
\end{equation}
respectively, in the $2+1$ decomposition discussed in Appendix~\ref{structure}
(compare~(\ref{2s}) and~(\ref{spinors})).  This matrix structure also makes
clear that $L$ conjugation maps $\Sp_\pm$ into each other.

We can decompose the Lie algebra $\so(5,1)$ as
\hbox{$\so(2)\oplus\so(3,1)\oplus(2\times\mb{4)}$},
interpreting $\so(3,1)$ as the Lorentz group, acting on $\{i,j,k,L\}$, and
where the last term refers to two vector representations of $\so(3,1)$,
labeled by $\{1,U\}$, each of which is 4-dimensional.
Since $\eS_L$ commutes with both $\so(3,1)$ and $\suR$, each eigenspace
$\Sp_\pm$ is also a representation of $\so(3,1)\oplus\suR$.  Since $\eX_1$
commutes with both $\so(3,1)$ and $\suR$, each eigenspace is in fact a
\textit{complex} representation of each of these algebras.  Although the
representations are reducible over any one of these algebras, each of
$\Sp_\pm$ is an irreducible representation ($\mb{4}_\CC$) over the sum.
(For convenience, we refer to these representations as ``pots''; see
Appendix~\ref{volume}.)

The Cartan basis can now be expanded to, say,
$\{\eX_1,\eS_L,\eD_{i,j},\eX_{kL},\eGS_k\}$, where it is important to notice
that the choice of Cartan elements in $\so(3,1)$ is independent of the choice
of Cartan element in $\suR$.  If we now interpret $\suR$ as the right-handed
weak symmetry algebra,
then each of these irreducible pots of spinors is an $\mb{8}$ containing a
(right-handed) weak doublet of Lorentz (Weyl) spinors.  
We can therefore interpret the eigenvalues of $\iota\circ\eD_{i,j}$ as spin
around the z axis and of $\iota\circ\eGS_k$ as (weak) particle type, and thus
identify spinor eigenstates of these Cartan elements with the Weyl spinors of
physical particles, as shown in Table~\ref{codeR} (compare Table~\ref{code} in
Appendix~\ref{binary}).  The (independent) identifications of particular
elements of a pot as having spin up or down, or as being an ``electron'' or
``neutrino'', depend on the Cartan elements chosen.

\begin{table}
\centering
\begin{tabular}{|c|c|c|}
\hline
\textbf{Type/Spin} & \boldmath $\Re(\bpsi)$, $\Im(\bpsi)$ & \textbf{Name} \\
\hline
&&\\[-11pt]
\hline
$++$ & $\eY_{1(\UU+L)}+\eZ_{k(\UU-L)}$, & $\nu_{\Ru}$ \\
& \quad $-\eY_{k(\UU+L)}-\eZ_{1(\UU-L)}$ & ($0$) \\
\hline
$+-$ & $\eY_{j(\UU+L)}+\eZ_{i(\UU-L)}$, & $\nu_{\Rd}$ \\
& \quad $-\eY_{i(\UU+L)}+\eZ_{j(\UU-L)}$ & ($0$) \\
\hline
$-+$ & $-\eY_{j(\UU+L)}+\eZ_{i(\UU-L)}$, & $e_{\Ru}$ \\
& \quad $-\eY_{i(\UU+L)}-\eZ_{j(\UU-L)}$ & ($-1$) \\
\hline
$--$ & $\eY_{1(\UU+L)}-\eZ_{k(\UU-L)}$, & $e_{\Rd}$ \\
& \quad $\eY_{k(\UU+L)}-\eZ_{1(\UU-L)}$ & ($-1$) \\
\hline
\end{tabular}
\caption{Half of the spinors in $\sl(3,\HH)$, showing one of the two complex
$\mb{4}$s of $\so(3,1)\oplus\so(4)$, along with their eigenvalues under
$\iota\circ\eGS_k$ and $\iota\circ\eD_{i,j}$ and their traditional name, with
their eigenvalue under hypercharge shown in parentheses.
}
\label{codeR}
\end{table}

\subsection{Interlude: Hypercharge}
\label{em}

We now have enough ingredients to introduce hypercharge.  There are four Weyl
spinors of $\so(3,1)$ in this theory, labeled by their eigenvalues under
$\eS_L$ ($\pm3$) and $\iota\circ\eGS_k$ ($\pm2$).
We can combine $\eS_L$ and $\eGS_k$ into the two-parameter family of operators
\begin{equation}
\HY_{gs} = s\,\eS_L + g\,\iota\circ\eGS_k
\end{equation}
with $s,g\in\RR$, where the presence of the complex structure $\iota$ is
required since $\eS_L$ has real eigenvalues whereas $\eGS_k$ has imaginary
eigenvalues.  Each operator $\HY_{gs}$ admits the four real eigenvalues
$\{-3s-2g,-3s+2g,3s-2g,3s+2g\}$, which are distinct so long as $3s\ne\pm2g$.
Assuming without loss of generality that $g,s>0$, there is a unique (up to
scale) operator $\HY$ that \textit{does} admit the degenerate eigenvalue $0$,
namely
\begin{equation}
\HY = h(2\,\eS_L + 3\,\iota\circ\eGS_k) .
\label{hyp}
\end{equation}

If $\HY$ does \textit{not} distinguish all four Weyl spinors,
we need a second operator to distinguish them.
The natural choice is the unique (again, up to scale) linear combination of
the two Cartan elements $\eS_L$ and $\eGS_k$ that is orthogonal to $\HY$ using
the Killing form, namely
\begin{equation}
\HYp = \iota\circ\eGS_k - \eS_L .
\end{equation}
We therefore use the eigenvalues of $\HYp$ to distinguish the four Weyl
spinors.
The operator $\HYp$ is orthogonal not only to $\HY$, but also to the Lorentz
group, $\so(3,1)$.  As noted in Section~\ref{e8}, $\HYp$ generates the
centralizer of $\su(5)$ in $\so(10)$.

Physically, we have identified a preferred Cartan element $\HY$ that admits
the degenerate eigenvalue $0$.
We identify $\HY$ as hypercharge, generating the (complex) $\uu_1$ of the
Standard Model; setting $h=-\frac{1}{12}$ will reproduce the correct values on
the eigenstates shown in the last column in Table~\ref{codeR}.

\subsection{\boldmath $\ee_{6(-26)}$}
\label{e6}

We now continue horizontally in the magic square from $\sl(3,\HH)$ to
$\ee_{6(-26)}$, henceforth referred to simply as $\ee_6$.  We add the
four remaining labels in $\OO$, thus expanding $\so(5,1)$ to $\so(9,1)$, as
discussed in more detail in~\cite{Lorentz}.  Thanks to triality, $\suR$ is
absorbed into the new $\so(4)$; its centralizer in $\so(4)$ is a second copy
of $\su(2)$, which we call $\suL$.  The algebra $\suL$ is in fact precisely
the subalgebra of $\gg_2$, the automorphism algebra of $\OO$, that fixes
$\HH$.  We have also expanded the Cartan basis by the addition of any single
element from $\suL$; although the choice is arbitrary, we choose $\eA_k$,
defined in Appendix~\ref{structure}, in order to simplify later calculations.

As well as adding a second $\su(2)$ subalgebra, 
going from $\sl(3,\HH)$ to $\ee_6$
also doubles the number of spinors, adding spinors labeled by
$\HH_\perp=\ell\HH$ to those labeled by~$\HH$.  
Not only does $\suL$ commute with the $\HH$-labeled spinors (since it
centralizes $\sl(3,\HH)$ in $\ee_6$), it also turns out that $\suR$ commutes
with the $\HH_\perp$-labeled spinors.

\begin{table}
\centering
\begin{tabular}{|c|c|c|}
\hline
\textbf{Type/Spin} & \boldmath $\Re(\bpsi)$, $\Im(\bpsi)$ & \textbf{Name} \\
\hline
&&\\[-11pt]
\hline
$++$ & $\eY_{k\ell(\UU+L)}+\eZ_{\ell(\UU-L)}$, & $\nu_{\Lu}$ \\
& \quad $-\eY_{\ell(\UU+L)}+\eZ_{k\ell(\UU-L)}$ & ($0$)\\
\hline
$+-$ & $\eY_{i\ell(\UU+L)}-\eZ_{j\ell(\UU-L)}$, & $\nu_{\Ld}$ \\
& \quad $\eY_{j\ell(\UU+L)}+\eZ_{i\ell(\UU-L)}$ & ($0$)\\
\hline
$-+$ & $\eY_{i\ell(\UU+L)}+\eZ_{j\ell(\UU-L)}$, & $e_{\Lu}$ \\
& \quad $-\eY_{j\ell(\UU+L)}+\eZ_{i\ell(\UU-L)}$ & ($-1$)\\
\hline
$--$ & $-\eY_{k\ell(\UU+L)}+\eZ_{\ell(\UU-L)}$, & $e_{\Ld}$ \\
& \quad $-\eY_{\ell(\UU+L)}-\eZ_{k\ell(\UU-L)}$ & ($-1$)\\
\hline
\end{tabular}
\caption{Half of the additional spinors in $\ee_6$, showing a complex $\mb{4}$
of $\so(3,1)\oplus\so(4)$, along with their eigenvalues under
$\iota\circ\eA_k$ and $\iota\circ\eD_{i,j}$ and their traditional name, with
their charge eigenvalue shown in parentheses.
}
\label{codeL}
\end{table}

Furthermore, the 32 spinor degrees of freedom in $\ee_6$ form two
Majorana--Weyl representations of $\so(9,1)$, again called $\Sp_\pm$,
distinguished as usual as eigenspaces of $\eS_L$, and labeled by $\UU\pm L$.
Elements of $\Sp_\pm$ are again Killing duals of each other.  Each such
eigenspace splits naturally into two ``pots'', each of which is an $\mb{8}$
that is a simultaneous complex representation of $\so(4)=\suR\oplus\suL$ and
$\so(3,1)$.  One pot is left-handed, consisting of doublets of $\suL$ and
singlets of $\suR$; the other is right-handed, consisting of doublets of
$\suR$ and singlets of $\suL$.  As before, the identification of (weak)
particle type and spin depend on the choice of Cartan elements.  However, the
pots themselves are independent of these choices, as discussed further in
Appendix~\ref{volume}.  We can again identify eigenstates of the Cartan
elements with physical particles, as shown in Table~\ref{codeL} (compare
Tables~\ref{codeR} and~\ref{code}).

We now have a theory containing a complex structure, $\so(3,1)$, $\suL$, and
$\suR$, as well as an additional $\so(1,1)$ generated by $\eS_L$, along with
\textit{two} sets of (complex!)\ spinors, each consisting of both a left- and
a right-handed weak doublet of Lorentz (Weyl) spinors.  As we will see in
Sections~\ref{DiracEq} and~\ref{particles}, suitable combinations of the
spinors in these two sets represent leptons and their antiparticles,
respectively.  Having introduced hypercharge in
Section~\ref{em}, and $\suL$ here, it is now straightforward to construct
the charge operator
\begin{equation}
q = -\frac16 \eS_L - \frac12 \eD_{i\ell,j\ell}
\label{charge}
\end{equation}
by combining hypercharge with the suitably-scaled Cartan element of $\suL$.
The resulting charge eigenvalues are shown in the last column in
Table~\ref{codeL}.

\subsection{\boldmath $\ee_{8(-24)}$}
\label{e8}

Our final stop is $\ee_8$ itself.  We add 
the labels in $\CC'_\perp$, thus expanding $\so(9,1)$ to $\so(12,4)$.
Thanks to triality, $\eS_L$ is absorbed in the new $\so(3,3)$, where it is
centralized by $\sl(3,\RR)$, which, as already noted in Section~\ref{sl3R}, is
a real form of complexified $\aa_2=\su(3)$.  We have added a color symmetry!
As discussed in Appendix~\ref{volume}, the single Majorana--Weyl spinor of
$\so(12,4)$ in $\ee_8$ splits into two $\mb{64}$s, denoted $\Sp_\pm$.  Each
of $\Sp_\pm$ is an irreducible representation of $\so(9,1)\oplus\so(3,3)$,
generalizing the language used in the previous sections.

We have thus added three new sets of ``colored'' spinors --
quarks~-- labeled by two new Cartan elements in $\sl(3,\RR)$, which we choose
to be $\eG_L$ and $\eA_L$ (see Appendix~\ref{structure}).
Since $\so(3,3)$ commutes with the complex structure $\eX_1$, its $\sl(3,\RR)$
subalgebra acts as \textit{complex} $\su(3)$ on these spinors, justifying the
name $\suc$ for this copy of $\sl(3,\RR)$.  The algebra $\suc$ is in fact
precisely the subalgebra of $\gg_2'$, the automorphism algebra of $\OO'$, that
fixes $L$, and hence $\CC'$.
(To the best of our knowledge, the first people to use this split form of
$\su(3)$ to describe the color symmetry of quarks were Gunaydin and
G\"ursey~\cite{Gunaydin73}.)

Apart from the action of $\suc$, colored spinors can be divided into
electroweak pots (representations of $\so(3,1)\oplus\so(4)$) exactly as we did for
the original, colorless pots.  However, the eigenvalues of $\eS_L$ are
different on colored spinors, $\pm1$ rather than $\pm3$, thus ensuring that
quarks have the correct fractional values of hypercharge and charge.

When moving to $\ee_8$, we have not merely added an $\so(3,3)$, but expanded
$\so(7,1)\subset\so(9,1)$ to $\so(10,4)\subset\so(12,4)$.  Since $\eX_1$
commutes with all of $\so(10,4)$, the spinors of $\ee_8$ are simultaneous
complex representations of Lorentz $\so(3,1)$ and the remaining $\so(7,3)$.
Thus, our theory contains all of the spinor pieces of the Georgi--Glashow
$\SO(10)$ GUT~\cite{GG}, along with its subtheories based on $\SU(5)$ and
$\SU(4)\times\SU(2)\times\SU(2)$, due to Georgi~\cite{GeorgiSO10} and
Pati--Salam~\cite{PS}, respectively.

Explicitly, this $\so(7,3)$ contains $\so(4)$, acting on labels in
$\HH_\perp$, as well as $\so(3,3)$, acting on labels in $\CC'_\perp$.  The
weak algebra $\so(4)$ of course splits as $\suR\oplus\suL$, where it is
noteworthy that $\suL\subset\gg_2$ (which $\suR$ is not), but that
$\suR\subset\sl(3,\HH)$ (which $\suL$ is not).  Similarly, $\so(3,3)$
contains both color $\sl(3,\RR)$ and $\eS_L$,
but $\sl(3,\RR)\subset\gg'_2$ (which $\{\eS_L\}$ is not), whereas
$\{\eS_L\}\subset\sl(3,\HH)$ (which this copy of $\sl(3,\RR)$ is not).

In this signature, $\so(7,3)$ itself does not contain any real form of
$\su(5)$.  Rather, on any spinor representation of $\so(7,3)$ we work with the
complexification of $\so(7,3)$ in the enveloping algebra, using $\iota$,
yielding (complex) $\so(10)$.  It is now straightforward to combine the
right-handed weak Cartan element (which commutes with $\suL$) with $\eS_L$
(which commutes with $\suc$) into the ``pure trace'' element $\HYp$
introduced in Section~\ref{em}.
A (complex) $\su(5)\subset\so(10)$ can now be defined as the subset of
$\so(10)$ that commutes with $\HYp$.  This $\su(5)$ will include not only
$\suL$ and $\suc$, but also the orthogonal linear combination $\HY$
introduced in Section~\ref{em}, which we recognize (when suitably normalized)
as the hypercharge operator of the $\su(5)$ GUT; $\iota\circ\HY$ generates the
$\uu_1$ of the Standard Model.

\section{Spinors and the Dirac Equation}
\label{DiracEq}

Having constructed $\ee_8$ in Section~\ref{construct} by traveling through the
magic square, we have seen how $\ee_8$ consists of $\so(12,4)$ together with a
single Majorana--Weyl spinor representation.  Along the way, we have
constructed $\so(12,4)$ from its Lorentz ($\so(3,1)$), weak
($\so(4)=\suR\oplus\suL$), and color ($\sl(3,\RR)\subset\so(3,3)$) sectors,
together with a complex structure given by the remaining $\so(2)$.  This
construction has allowed us to interpret the 128 spinor degrees of freedom as
complex Weyl spinors of $\so(3,1)$ and their Killing duals.  How do they combine
to form the Dirac spinors used to represent particles and antiparticles?

We begin with the lepton sector, with labels in \hbox{$\CC'\otimes\OO$}, which
lives in $\ee_6\subset\ee_8$.
The $2\times2$ subalgebra of $\ee_6$ is $\so(9,1)$, which decomposes into the
complex structure, $\so(7,1)$ and two vector $\mb{8}$s.  Either of these
vector representations can be used to generate $\Cl(7,1)$ in the enveloping
algebra of $\so(9,1)$, acting on spinors; $\Cl(7,1)$, of course, contains the
Lorentz Clifford algebra, $\Cl(3,1)$.  The chosen representation can then be
interpreted as the degree-1 elements (gamma matrices) of $\Cl(7,1)$.  Since
the commutator of the two representations is proportional to the complex
structure, which is also the volume element $\Omega_{7,1}$ (see
Appendix~\ref{volume}), the other representation corresponds to
\textit{pseudovectors}, that is, degree-7 elements of $\Cl(7,1)$.  We make the
arbitrary choice to generate $\Cl(7,1)$ with the vector representation that
contains the energy basis element $D_L$.

As discussed in Appendix~\ref{volume}, the lepton spinors divide naturally
into two irreducible representations $\Sp_\pm$ of $\so(9,1)$.  Each
of these subspaces is totally null under the Killing form $B$, and there is a
natural pairing between them given by $L$ conjugation ($\vp$), defined in the
enveloping algebra in Appendix~\ref{envelop}.

As we proposed in~\cite{Spin}, we interpret the additional spacelike
dimensions as possible masses, thus converting the massive Dirac equation in
four spacetime dimensions to the massless Weyl equation in higher dimensions.
So we seek spinor solutions $\psi$ of the equation
\begin{equation}
[Q,\psi] = 0
\label{DiracQ}
\end{equation}
where $Q$ lies in our chosen vector representation of $\so(7,1)$, and
therefore has the form
\begin{equation}
Q = E \eD_L + p_a \eX_a + m_b \eX_b
\label{Qmom}
\end{equation}
with $a=i,j,k\in\Im\HH$ and $b=k\ell,j\ell,i\ell,\ell\in\HH_\perp$.  We
further write $Q=P+M$, where $P$ is the Lorentz vector, consisting of the first
four terms of $Q$, and $M$ is the mass vector, consisting of the remaining
four terms.

Recall that each of $\Sp_\pm$ divides naturally into a right-handed pot with
labels in $\HH$ (and hence inside $\aa_5$) and a left-handed pot with labels
involving $\ell$.  Since $P\in\aa_5$ and $M\not\in\aa_5$, it is clear that $P$
preserves pots whereas $M$ maps between pots.  Writing $\psi=\psi_L+\psi_R$,
(\ref{DiracQ}) becomes
\begin{align}
[P,\psi_L] + [M,\psi_R] &= 0 ,\label{DiracL}\\
[P,\psi_R] + [M,\psi_L] &= 0 .\label{DiracR}
\end{align}
Since $Q$ is a degree-1 element of $\Cl(7,1)$, we can use the Clifford
identity~(\ref{Clifford})
to solve~(\ref{DiracL}) for $\psi_L$ in terms of $\psi_R$, obtaining
\begin{equation}
|P|^2 \psi_L = +\bigl[P,[M,\psi_R]\bigr]
\label{DiracLR}
\end{equation}
which we can then substitute into~(\ref{DiracR}), resulting in
\begin{equation}
|P|^2\,[P,\psi_R] + \Bigl[M,\bigl[P,[M,\psi_R]\bigr]\Bigr] = 0  .
\end{equation}
Since $P$ and $M$ anticommute, and once again using the Clifford identity, we
have
\begin{equation}
\Bigl[M,\bigl[P,[M,\psi_R]\bigr]\Bigr]
  = -\Bigl[M,\bigl[M,[P,\psi_R]\bigr]\Bigr]
  = |M|^2 [P,\psi_R]
\end{equation}
and thus finally obtain
\begin{equation}
|Q|^2 = |P|^2+|M|^2 = 0
\label{Qnorm}
\end{equation}
as expected.  That is, solutions to the eight-dimensional Weyl
equation~(\ref{DiracQ}) only exist if $Q$ is null.  Furthermore, for given
$Q$, \textit{any} right-handed spinor $\psi_R$ determines a unique left-handed
partner $\psi_L$, given by~(\ref{DiracL}), such that the resulting Dirac
spinor $\psi$ satisfies~(\ref{DiracQ}).  We henceforth refer to~(\ref{DiracQ})
as the generalized Dirac equation, and its solutions as Dirac spinors.

This construction can be reversed; we can recover the 8-momentum $Q$ (up to
scale) from Dirac spinors as the vector bilinear
\begin{equation}
Q_\psi = \bigl[[\vp(\psi),\eD_L],\psi\bigr] .
\label{bilin}
\end{equation}
The generalized Dirac equation~(\ref{DiracQ}) is thus a special case of the
``3-$\psi$ rule''~\cite{FairlieII,Sudbery,Schray,NonReal}, an identity on
(Dirac) spinors.
Thanks to the Jacobi identity, both~(\ref{DiracQ}) and~(\ref{bilin}) are
covariant under $\so(12,4)$, and in fact under all of $\ee_8$.

We have already labeled (four-dimensional) Weyl spinors by their eigenvalues
with respect to our chosen Cartan basis; see Tables~\ref{codeR},~\ref{codeL}
and~\ref{code}.  These eigenvalues encode physical properties such as spin and
particle type.  We seek Dirac spinors that combine two Weyl spinors with the
same physical properties.
Our ability to achieve this goal is constrained by~(\ref{DiracLR}).  If
$\psi_R$ is an eigenvector of a particular Cartan element, does $\psi_L$ have
the same eigenvalue?

We must first deal with the potential complication that the complex structure
\textit{anticommutes} with $Q$, that is, with both $P$ and $M$.
%
%
Nonetheless, the complex structure commutes with the combination $P\circ M$
that appears in~(\ref{DiracLR}), thus ensuring that solutions of the Dirac
equation are complex, that is, if $\psi$ satisfies~(\ref{DiracQ}) for given
$Q$, so does $\iota(\psi)$.

Turning next to our spin Cartan element, which we have chosen to be
$L_z=\iota\circ\eD_{i,j}$, we discover that $L_z$ only commutes with
$P\circ M$ if the momenta $p_a$ vanish for $a=i,j$.  This is the expected
result: the eigenvalue of $L_z$ only corresponds to spin (or helicity) if the
$x$- and $y$-momenta vanish.

We finally consider particle type.  Since each chiral $\su(2)$ acts only on one
of the Weyl spinors $\psi_L$ and $\psi_R$, we must add the two Cartan elements
to get an operator whose eigenvalues will be the same for both of these
spinors, a property that we have already used in~(\ref{charge}).  As discussed
more fully in Section~\ref{generations}, we have chosen these Cartan elements
so that the diagonal subalgebra of $\so(4)$ obtained by adding corresponding
elements of $\suR$ and $\suL$ is the $\so3$ that fixes $\ell$ in $\so(4)$.
With this choice, the combined weak Cartan element which forms part of the
charge operator, is $\eW_z=\iota\circ\eD_{i\ell,j\ell}$.  Similarly to the
analysis above for spin, in order for a Dirac spinor,
satisfying~(\ref{DiracLR}), to have a well-defined weak eigenvalue, that is,
be an eigenstate of $\eW_z$, two of the mass components must vanish, namely
$m_b=0$ for $b=i\ell,j\ell$.

Extending the above description of leptons in $\ee_6$ to quarks in $\ee_8$ is
straightforward.  As pointed out in Appendix~\ref{volume}, the spinor
eigenspaces $\Sp_\pm$ are now irreducible representations of
$\so(9,1)\oplus\so(3,3)$, and we obtain colored versions of the spinor eigenstates
in Tables~\ref{codeR} and~\ref{codeL}, with null labels $I\pm IL$, $J\pm JL$,
$K\pm KL$ instead of $1\pm L$.  (The correct signs can be determined by
commuting the lepton spinors with elements of $\so(3,3)$ such as $\eD_{J,K}$.)
As discussed in Appendix~\ref{code}, three additional digits can be added to
the binary code to account for color.

\section{Particles and Antiparticles}
\label{particles}

For massive particles, the eigenvalues of the Cartan element $L_z$ correspond
to spin \textit{at rest}, that is, with $p_k=0$ (in addition to the
assumptions above).  Using our principle that $\ell$ should be ``special,'' we
will also make the simplifying assumption that $m_{k\ell}=0$.
Tables~\ref{codeR}, \ref{codeL} and~\ref{code} were constructed with these
conventions in mind; when combined into Dirac spinors by addition across
tables, all of the resulting particles are at rest, and the only nonzero mass
is $m_\ell$.  A discussion of the more general case appears in
Section~\ref{discuss}.

At rest, our general 8-momentum~(\ref{Qmom}) has been reduced to
\begin{equation}
Q = E \eD_L + m \eX_\ell
\label{Q5}
\end{equation}
and we arbitrarily assume that $E,m>0$ corresponds to \textit{particles}.  The
Dirac equation for a particle with mass $m$ and energy $E$ is therefore
\begin{equation}
[E \eD_L + m \eX_\ell, \psi] = 0
\label{Dirac}
\end{equation}
with $E^2=m^2$ following from~(\ref{Qnorm}).  What equation does the Killing
dual $\vp(\psi)$ satisfy?  The operation $\vp$ commutes with the spatial
momenta and masses $\eX_q$ for $q\in\Im\OO$, but \textit{anticommutes} with
the energy $\eD_L$, so we have
\begin{equation}
[- E \eD_L + m \eX_\ell, \psi] = 0
\end{equation}
that is, $\psid$ satisfies a Dirac equation with negative energy.  What about
its charge?  The hypercharge operator defined in Section~\ref{em} contains
$S_L$, which \textit{anticommutes} with $\vp$, so the possible charges for
$\psid$ are opposite those of $\psi$.  However, both the spin Cartan element
$L_z$ and the combined weak Cartan element $\eW_z$ \textit{commute} with $\vp$,
so $\psid$ doesn't quite have the opposite charge from $\psi$.

These issues with the mass and charge can both be resolved at once by defining
\begin{equation}
\psib = [\vp(\psi),\eD_L]
\end{equation}
since $\eD_L$ anticommutes with both $\vp$ and $\eX_\ell$.  Thus, $\psib$
satisfies the same (uncoupled) Dirac equation as $\psi$, namely~(\ref{Dirac}),
but has the opposite charge.  We therefore identify $\psib$ as the
\textit{antiparticle} of $\psi$.  Thus, particles live in $\Sp_+$, and
antiparticles live in $\Sp_-$.  However, negative energy solutions
of~(\ref{DiracQ}) also exist in $\Sp_\pm$, namely the Killing duals of
antiparticles and particles, respectively.
We reiterate that, as pointed out in Section~\ref{sl3R}, we consider Killing
duals to represent adjoints and not separate physical degrees of freedom.

We can perform a Lorentz transformation on solutions of the Dirac equation
at rest to obtain particle solutions of the Dirac equation with arbitrary
momentum.

We can now finally interpret the two spinor spaces $\Sp_\pm$ first introduced
in Section~\ref{sl3R} and generalized in Appendix~\ref{volume}.  As discussed
in Section~\ref{em}, elements of these spaces have opposite eigenvalues under
$\eS_L$, and hence opposite charges.  By the reasoning just given above, each
of these spaces contains both particles and antiparticles.  We have shown that
$\Sp_+$ contains particles and the Killing duals of antiparticles, while the
antiparticles themselves live in $\Sp_-$, along with the Killing duals of
particles, as illustrated in Table~\ref{codeP}.  As this table makes clear,
the traditional names used in Tables~\ref{codeR} and~\ref{codeL} for Weyl
spinors are ambiguous until they are combined into Dirac spinors with the
correct relative sign.

\begin{table}
\centering
\begin{tabular}{|c|c|c|}
\hline
\textbf{Type/Spin} & \textbf{Dirac Spinor} & \textbf{Name} \\
\hline
&&\\[-11pt]
\hline
$++$ & $\nu_{\Lu}+\nu_{\Ru}\in\Sp_+$ & $\nu_\up$ ($0$) \\
\hline
$++$ & $\nu_{\Lu}-\nu_{\Ru}\in\Sp_+$ & $\vp(a(e_\up))$ ($0$) \\
\hline
$++$ & $\vp(\nu_{\Lu})-\vp(\nu_{\Ru})\in\Sp_-$ & $a(e_\up)$ ($1$) \\
\hline
$++$ & $\vp(\nu_{\Lu})+\vp(\nu_{\Ru})\in\Sp_-$ & $\vp(\nu_\up)$ ($1$) \\
\hline
$-+$ & $e_{\Lu}+e_{\Ru}\in\Sp_+$ & $e_\up$ ($-1$) \\
\hline
$-+$ & $e_{\Lu}-e_{\Ru}\in\Sp_+$ & $\vp(a(\nu_\up))$ ($-1$) \\
\hline
$-+$ & $\vp(e_{\Lu})-\vp(e_{\Ru})\in\Sp_-$ & $a(\nu_\up)$ ($0$) \\
\hline
$-+$ & $\vp(e_{\Lu})+\vp(e_{\Ru})\in\Sp_-$ & $\vp(e_\up)$ ($0$) \\
\hline
\end{tabular}
\caption{Combinations of spinors from Tables~\ref{codeR} and~\ref{codeL} that
correspond to particles ($\psi$), antiparticles ($a(\psi)$), and their Killing
duals (using $\vp$), with charge eigenvalues (using~(\ref{charge})) in
parentheses.  Only spin-up eigenstates are shown; the spin-down case is
similar.}
\label{codeP}
\end{table}

\section{Decompositions and Mediators}
\label{mediators}

We have chosen to construct $\ee_8$ from smaller Lie algebras, as we feel that
the construction process most clearly demonstrates the fundamental nature
played by each constituent.  Our preferred path, outlined in
Section~\ref{construct}, introduces the complex structure first, then the
timelike coordinate $L$, then both the Lorentz algebra $\so(3,1)$ and the
right-handed weak algebra $\suR$, and only then the left-handed weak algebra
$\suL$ followed by color $\suc$.  This process can of course be reversed, by
starting with $\ee_8$ and decomposing it into smaller Lie algebras together
with their representations.

The first step in this decomposition is to choose a preferred copy of
$\so(12,4)$ in $\ee_8$; as outlined in Appendix~\ref{structure}, this step
amounts to choosing a preferred $2\times2$ block structure in the $3\times3$
matrix representation.  The remaining $248-120=128$ elements form a single
Majorana--Weyl representation of $\so(12,4)$, thus dividing $\ee_8$ into
``adjoint'' and ``spinor'' sectors.

This remarkable division of an exceptional Lie algebra into adjoint and spinor
representations of a smaller Lie algebra is in fact a unifying feature of all
of the Lie algebras in the magic square.  In each case, the Cartan subalgebra
of the maximal subalgebra (as given in Table~\ref{subalg}) is also the Cartan
subalgebra of the full ($3\times3$) Lie algebra.  It is well known that any
(complex) simple Lie algebra can be expressed in terms of its chosen Cartan
subalgebra and its eigenstates, which serve as raising and lowering operators
for any Cartan basis.  Since the maximal subalgebra and the full algebra have
the same Cartan subalgebra, we obtain three types of elements: the Cartan
subalgebra, its simultaneous eigenstates in the maximal subalgebra, and the
remaining simultaneous eigenstates in the $3\times3$ block.  The latter
elements are spinor eigenstates of the chosen Cartan subalgebra, on which the
eigenstates in the maximal subalgebra act as raising and lowering operators.

There is one further subtlety: constructing eigenstates requires a complex
structure, and our Lie algebras are real.  However, we can interpret one of
the ``rotation-like'' Cartan elements as a complex structure, thus
complexifying the spinor representation at the cost of a slight reduction in
the size of the adjoint piece.  The remaining elements of the original adjoint
can be identified as the degree-$1$ and degree-($n-1$) elements of $\Cl(n)$, as
discussed in Appendix~\ref{volume}.

We therefore decompose $\so(12,4)$ as
$\so(10,4)\oplus\so(2)\oplus(2\times\mb{14})$.
The $\so(2)$ acts as a complex structure for the action of $\so(10,4)$ on
the $\mb{128}$ spinor representation, realizing it as a single Weyl spinor
representation of $\so(10,4)$, that is, $\mb{64}_\CC$.  We have therefore
decomposed the real Lie algebra $\ee_{8(-24)}$ into \textit{complex}
$\so(14)$, a single Weyl spinor of $\so(14)$, and a complex 14-vector 
that can be used to represent the degree-$1$ and degree-$13$ elements of
$\Cl(10,4)$.

The stage is now set to decompose complex $\so(14)$ (in the enveloping
algebra) into the desired physical symmetry algebras, and the spinors into
representations of those algebras.  As outlined in Section~\ref{construct},
only in reverse, $\so(10,4)$ contains Lorentz $\so(3,1)$ together with
$\so(7,3)$; the latter, when complexified, yields $\so(10)$, leading directly
to the Georgi--Glashow $\so(10)$ GUT~\cite{GG}, along with its subalgebras
$\su(5)$ and $\su(4)\oplus\su(2)\oplus\su(2)$, and hence their corresponding
GUTs, due to Georgi~\cite{GeorgiSO10} and Pati--Salam~\cite{PS}, respectively.

Within $\ee_6$, our spinor pots are irreducible representations of
$\so(3,1)\oplus\so(4)\subset\so(7,1)$.  What role is played by the remaining 16
elements of $\so(7,1)$?  These elements comprise four Lorentz vectors, with
weak labels $i\ell$, $j\ell$, $k\ell$, $\ell$, suggesting their interpretation 
as the electroweak
gauge potentials.

Since the complex structure $\iota$ commutes with $\so(7,1)$, we can find an
eigenbasis of this $4\times\mb{4}$, that is, a basis of simultaneous raising
and lowering operators for the four Cartan elements of $\so(3,1)\oplus\so(4)$
(when acting on spinors).  Half of these basis elements commute with spin
($L_z$); the other half raise or lower the spin eigenvalue.  Similarly half
commute with the (combined) weak Cartan element $\eW_z$; the other half change
the weak eigenvalue and hence the particle type, as expected.  We are puzzled
by the fact that, since each element of the $4\times\mb{4}$ necessarily
anticommutes with the volume elements $\omega$ and $\Omega$ of $\so(3,1)$ and
$\so(4)$, that is, with $\gamma_5$, these weak mediators mix right- and
left-handed spinors.

Similarly, when adding color, we extended $\so(9,1)$ to
$\so(9,1)\oplus\so(3,3)$, thus also adding, among other things, six (not 8!)
new vectors of $\so(3,1)$.  Applying an analogous analysis leads to six
possible gauge potentials, with particular behavior regarding spin and color
eigenvalues when acting on spinors.

If, as suggested by the structure of $\ee_8$, these $4+6=10$ Lorentz vectors
are indeed mediators, we are led to weak mediators that form a vector $\mb{3}$
of the combined weak algebra $\so(3)$, and to color mediators that form a
$\mb{3}\oplus\bar{\mb{3}}$ of $\suc$, rather than the adjoint representations
predicted by the Standard Model.  

\section{Generations}
\label{generations}

The decomposition of $\so(4)$ into two copies of $\su(2)$ is well known, as is
the fact that adding appropriate commuting elements of these subalgebras
yields the rotation subalgebra $\so(3)$.  However, an underappreciated
property of this decomposition is that the two $\su(2)$ subalgebras are in
fact uniquely determined, which follows from the uniqueness of the
decomposition of semisimple Lie groups into commuting simple Lie groups.  In
particular, the $\su(2)$ subalgebras do \textit{not} depend on a choice of
$\so(3)$ subalgebra.  Which $\so(3)$ subalgebra one gets depends on the chosen
correspondence between the two $\su(2)$ subalgebras.

Without loss of generality, we chose $\ell\in\HH_\perp$ to be the axis fixed
by $\so(3)$.  However, even with the pairing fixed between elements of
$\suL$ and $\suR$, we still have the choice of which pair to choose as the
basis of the Cartan subalgebra of $\so(4)$.  We chose $\{\eA_k,\eGS_k\}$
above, but we could equally well have replaced $k$ with $i$ or $j$.
Equivalently, we have the choice of Cartan element in $\so(3)$; the choice made
above is $\eD_{i\ell,j\ell}$, where we could cycle $i\ell$, $j\ell$, $k\ell$.

What effect would such a change in the Cartan element have on the particle
eigenstates?  In quantum mechanics, it is well understood how to combine spin
eigenstates along one axis in order to get spin eigenstates along another,
corresponding to a different choice of angular momentum Cartan element.  Here,
we have the analogous ability to combine weak eigenstates of one ``type'' in
order to get weak eigenstates of another ``type,'' corresponding to a
different choice of weak Cartan element.  We propose interpreting these
different types as \textit{generations}.

In this interpretation, the generations sit on top of each other, in the sense
that they belong to a single pot.  However, each generation corresponds to a
particular choice of Cartan elements in $\so(4)$, leading to a particular
division of the pots into Weyl spinors of $\so(3,1)$, representing the weak
eigenstates of that generation.

Is this generation structure compatible with observation?  Answering that
question would require a mechanism to break the continuous symmetry,
presumably also accounting for the different masses associated with each
generation.  Since the charge operator~(\ref{charge}) incorporates the
$\so(3)$ Cartan element, it would be generation dependent, although the
resulting charge eigenvalues would be the same.  Just as there are three spin
axes, this model has three natural ``generation'' axes.  Furthermore, the
overlapping nature of the generations in this model suggests a natural
framework for the observed mixing between generations.

\section{Chirality}
\label{chiral}


Although there are several notions of ``chiral'' theories in the literature,
the real question is whether a theory describes the chiral asymmetry seen in
nature.  Since the theory presented here does not (yet) describe
interactions, that question can not (yet) be answered directly.

\citet{Distler} define a chiral $E_8$ theory to be, essentially, one in which
the spinors do not have a self-conjugate structure, then argue that no such
theory exists.  However, they assume both that the GUT group is compact, and
that $\ee_8$ has been complexified, neither of which holds for our model.


Interestingly, there are several senses in which our model is fundamentally
``chiral.''  First of all, $\ee_{8(-24)}$ contains a single Weyl spinor of
$\so(10,4)$.  As a consequence, the seven ``bits'' of information in the
spinor binary code are not independent, as discussed in Appendix~\ref{code}.
Thus, ``Lorentz handedness'' (eigenvalue of the $\so(3,1)$ volume element
$\omega$) and ``weak handedness'' (eigenvalue of the $\so(4)$ volume element
$\Omega$) are correlated.
If we were to complexify $\ee_8$, as required by~\cite{Distler}, then this
complexified $\ee_8$ would contain spinors of both $\so(10,4)$ handednesses.
We reiterate, however, that we work throughout with a real form of $\ee_8$.

Furthermore, our route through the magic square has a stop at $\aa_5$, which
contains $\suR$ but not $\suL$.  Even though these algebras appear
symmetrically as $\so(4)\subset\so(12,4)$, that symmetry is misleading, as
$\suL$, but not $\suR$, is in $\gg_2$, as was pointed out in Section~\ref{e8}.
Is this asymmetry enough to result in a chiral theory?

\section{Conclusion}
\label{conclude}

We cannot overemphasize the extent to which our construction is driven by the
underlying mathematical structure.  We make two key assumptions:
\begin{itemize}
\item
We work with a \textit{real} Lie algebra, $\ee_{8(-24)}$, \textit{without}
complexifying it.
\item
\textit{All} objects in the theory are constructed from the same copy of
$\ee_{8(-24)}$.
\end{itemize}
We find not only the Standard Model symmetry algebras, including Lorentz
symmetry, as well as the appropriate notion of complexification, but also the
spinors on which they act, their momenta, and possibly the mediators and a
version of the Higgs particle, all within a particular real form of $\ee_8$.
We refer to the particles among these objects collectively as \textit{octions}.
Furthermore, as emphasized by our chosen path through the magic square, by
taking the division algebra structure seriously we see that the $\so(4)$ and
$\so(3,3)$ subalgebras of $\so(12,4)$ break \textit{naturally} into
$\suL\oplus\suR$, and $\so(1,1)\oplus\sl(3,\RR)$, respectively.
Recall from Table~\ref{subalg} that the centralizer of $\sl(3,\HH)$ is
precisely $\suL\oplus\sl(3,\RR)$, corresponding to the weak and strong
interactions.  Thus, the unitary groups in the standard model are a natural
feature of our construction, even though triality later reveals them to be
subalgebras of the orthogonal algebra $\so(12,4)$.


We reiterate that the action of $\so(7,3)\oplus\so(3,1)\subset\so(12,4)$ on
spinors is complex, using the complex structure in the enveloping algebra.
In particular, $\so(7,3)$ acts as $\so(10)$, and $\sl(3,\RR)\subset\so(3,3)$
acts as color $\su(3)$ on spinors as usual.

It is remarkable that $\ee_8$, the largest exceptional Lie algebra, turns out
to be precisely the right size to capture the essential content of the
Standard Model.  There is no wasted space: Using two of the 16 degrees of
freedom for the complex structure and 10 for the internal symmetries of the
Standard Model \textit{requires} the external world to be 4-dimensional, and
described by special relativity.  Furthermore, as we have shown, the 128
spinor degrees of freedom correspond precisely to one generation of leptons
and quarks (and their Killing adjoints).  We have also described how this
description naturally generalizes to three overlapping generations.

For several decades, a number of authors have been exploring the relationship
between division algebras and the Standard Model and, more recently, GUTs
based on $\so(10)$ in ways that parallel our identification here of $\so(7,3)$
inside of $\ee_8$.  We believe that these authors have been seeing the same
physical structures that we have described, but without the guidance of our
two key assumptions.

\section{Future Work}
\label{discuss}

We speculate here on several unanswered questions that we hope to address in
the future.

\begin{enumerate}

\item
It would be immensely helpful to have an action for this theory.  The spinor
interaction terms in the Standard Model action involve contractions between
elements of spinor and vector representations of $\so(3,1)$ and elements of
$\Cl(3,1)$.  Working in momentum space, all of these objects can be
represented in the enveloping algebra of $\ee_8$, with the Killing form
playing the role of the (real part of the) contraction.  But a straightforward
translation of the Standard Model action in this way turns out to be
cumbersome and unilluminating.

The more interesting question is whether the result can be written nicely in
terms of our division algebra description of $\ee_8$.  Preliminary
calculations suggest that this is at least partially the case, with $\so(7,1)$
momenta playing the roles of $\so(3,1)$ momenta and mass, and a sum over pots
 as a sum both over particle types and over generations.  To complete
this program, it is necessary to better understand the proposed electroweak
mediators (described in Section~\ref{mediators}) that always link right handed
spinors to left-handed ones.  Is such a theory chiral, and if so, why?

\item
In the absence of a full, interacting theory, it is difficult to discuss
whether that theory will satisfy the requirements of the Coleman--Mandula
theorem~\cite{CM,Mandula15}.  For instance, the theorem assumes Poincar\'e
invariance of massive particles, but the Weyl spinors in our unbroken theory
are massless.  By the time the symmetry is broken so that (massive) Dirac
spinors can be identified, the symmetry group has been reduced to a direct
product of the Lorentz group and the internal symmetry group, as the theorem
demands.
Ultimately, we will need an action that has this direct product as its
symmetry group.

\item  
Does the prediction of only six gluons, transforming as a $\mb{3}$ and
$\mb{\bar{3}}$ of $\sl(3,\RR)$ agree with experiment?

\item
It is possible to add dimensions and scales to a Lie algebra by multiplying
the Cartan elements by dimensionful constants.  Can this be done in a
consistent way within $\ee_8$ that explains why the fundamental constants of
nature are what they are?

\item
In Section~\ref{particles}, we assumed that the only nonzero mass was
$m_\ell$.  For completeness, we note here that choosing $m_{k\ell}\ne0$ still
preserves weak eigenvalues $\psi_L$ and $\psi_R$ under the action of a weak
raising or lowering operator.  However, the resulting Dirac spinor satisfies a
slightly different Dirac equation, namely one in which the sign of $m_{k\ell}$
has flipped.  Can such a mechanism be used, for instance, to provide different
masses to electrons and neutrinos?

\item
Can the description of the three overlapping generations be used to explain
the role of the CKM (quark) and PMNS (lepton) mixing matrices in the Standard
Model?

\item
When breaking $\so(12,4)$ to $\so(10,4)$ in Section~\ref{mediators}, the
degree-$1$ elements of $\Cl(3,1)\subset\Cl(10,4)$ were interpreted, in
Section~\ref{DiracEq}, as the energy and momentum of the fermions.  The
$2\times10$ elements of $\Cl(10,4)$ that are simultaneously Lorentz scalars
and a vector representation of the internal symmetries suggest an
identification as generalized Higgs degrees of freedom.  In
Section~\ref{DiracEq}, when we broke weak $\so(4)$ into $\so(3)$ by choosing
$\ell$ to be special, we added the corresponding degree-$1$ element of the
Clifford algebra to the energy/momentum degrees of freedom in order to provide
a mass for the fermions.  Could this process be used to describe the usual
Higgs symmetry breaking?

\item
The canonical commutation relations between position and momentum closely
parallels that between vectors and pseudovectors as discussed in
Appendix~\ref{volume}.  We have used the degree-$1$ elements of $\Cl(10,4)$ to
describe (generalized) momenta.  Could the degree-$13$ elements of $\Cl(10,4)$
corresponding to Lorentz $\so(3,1)$, which are also degree-$3$ elements of
$\Cl(3,1)$, represent physical spacetime?

\end{enumerate}

\goodbreak
\acknowledgments{%
CAM and TD would like to thank David Fairlie and Tony Sudbery for introducing
them to the octonions and their possible uses in particle physics, J\"org
Schray for his work with us on the transition from octonionic Lie algebras to
Lie groups, Aaron Wangberg for elucidating the structure of~$\ee_6$, and Jim
Wheeler for teaching them about the conformal group.  They also thank the
participants in the series of Octoshop workshops, whose many insights into the
relationship between octonions and physics are deeply and inextricably
embedded in this work.

RAW would like to thank John Conway for introducing him to the octonions and
their uses in group theory

This work was supported in part by the John Templeton Foundation under grant
number 34808, by FQXi, and by the Institute for Advanced Study.
}

\goodbreak
\appendix

\begin{figure}
\centering
\includegraphics[width=5cm]{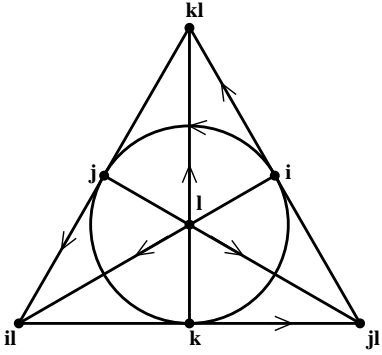}
\caption{A graphical representation of the  octonionic multiplication table.}
\label{omult3}
\end{figure}

\begin{table}
\begin{tabular}[b]{|c|c|c|c|c|c|c|c|}
\hline
&\boldmath$i$&\boldmath$j$&\boldmath$k$&\boldmath$k\ell$
  &\boldmath$j\ell$&\boldmath$i\ell$&\boldmath$\ell$\\\hline
\boldmath$i$&$-1$&$k$&$-j$&$j\ell$&$-k\ell$&$\ell$&$i\ell$\\\hline
\boldmath$j$&$-k$&$-1$&$i$&$-i\ell$&$\ell$&$k\ell$&$j\ell$\\\hline
\boldmath$k$&$j$&$-i$&$-1$&$-\ell$&$i\ell$&$-j\ell$&$k\ell$\\\hline
\boldmath$k\ell$&$-j\ell$&$i\ell$&$\ell$&$-1$&$i$&$-j$&$-k$\\\hline
\boldmath$j\ell$&$k\ell$&$\ell$&$-i\ell$&$-i$&$-1$&$k$&$-j$\\\hline
\boldmath$i\ell$&$\ell$&$-k\ell$&$j\ell$&$j$&$-k$&$-1$&$-i$\\\hline
\boldmath$\ell$&$-i\ell$&$-j\ell$&$-k\ell$&$k$&$j$&$i$&$-1$\\\hline
\noalign{\vspace{0.15in}}
\end{tabular}
\caption{The octonionic multiplication table.}
\label{omult}
\end{table}

\begin{table}
\centering
\small
\begin{tabular}{|c|c|c|c|c|c|c|c|}
\hline
$$&\boldmath$I$&\boldmath$J$&\boldmath$K$
  &\boldmath$KL$&\boldmath$JL$&\boldmath$IL$&\boldmath$L$\\\hline
\boldmath$I$&$-1$&$K$&$-J$&$JL$&$-KL$&$-L$&$IL$\\\hline
\boldmath$J$&$-K$&$-1$&$I$&$-IL$&$-L$&$KL$&$JL$\\\hline
\boldmath$K$&$J$&$-I$&$-1$&$-L$&$IL$&$-JL$&$KL$\\\hline
\boldmath$KL$&$-JL$&$IL$&$L$&$1$&$-I$&$J$&$K$\\\hline
\boldmath$JL$&$KL$&$L$&$-IL$&$I$&$1$&$-K$&$J$\\\hline
\boldmath$IL$&$L$&$-KL$&$JL$&$-J$&$K$&$1$&$I$\\\hline
\boldmath$L$&$-IL$&$-JL$&$-KL$&$-K$&$-J$&$-I$&$1$\\\hline
\end{tabular}
\caption{The split octonionic multiplication table.}
\label{smult}
\end{table}

\section{\boldmath The Structure of $\ee_8$}
\label{structure}

We summarize here the construction of the adjoint representation of $\ee_8$
given in~\cite{Magic}.  We work throughout with the ``half-split'' real form
$\ee_{8(-24)}$, referred to throughout as $\ee_8$, although only minor changes
are required to handle the other real forms.  The \textit{octonions} $\OO$ are
the real algebra spanned by the identity element~$1$ and seven square roots
of~$-1$ that we denote $i$, $j$, $k$, $k\ell$, $j\ell$, $i\ell$, $\ell$, whose
multiplication table is neatly described by the oriented Fano geometry shown
in Figure~\ref{omult3}, and given explicitly in Table~\ref{omult}.  The
\textit{split octonions} $\OO'$ are the real algebra spanned by the identity
element~$U$, three square roots of~$-1$ that we denote $I$, $J$, $K$, and four
square roots of~$+1$ that we denote $KL$, $JL$, $IL$, $L$, whose
multiplication table is given in Table~\ref{smult}.  We will often write $U$
as~$1$ when there is no ambiguity, as in Table~\ref{smult}.

Although there are many complex and quaternionic subalgebras of $\OO$ and
$\OO'$, we normally take the quaternions $\HH$ to be spanned by $\{1,i,j,k\}$,
and the split complex numbers $\CC'$ to be spanned by $\{U,L\}$.  The
orthogonal complements of these algebras will be denoted as
$\HH_\perp\subset\OO$ and $\CC'_\perp\subset\OO'$, respectively.

As shown in~\cite{Magic}, almost all of $\ee_8$ can be represented as
$3\times3$ antihermitian tracefree matrices over $\OO'\otimes\OO$, thus
justifying the alternate name $\su(3,\OO'\otimes\OO)$ for $\ee_8$.  There are
$3\times8\times8+2\times(7+7)=220$ independent such matrices, 
which take the form
\begin{align}
\eD_q &= \begin{pmatrix}q& 0& 0\\ 0& -q& 0\\ 0& 0& 0\\\end{pmatrix} ,
\quad
\eX_p  = \begin{pmatrix}0& p& 0\\ -\bar{p}& 0 &0\\ 0& 0& 0\\\end{pmatrix} ,
\nonumber\\
\eY_p &= \begin{pmatrix}0& 0& 0\\ 0& 0& p\\ 0& -\bar{p} &0\\\end{pmatrix} ,
\quad
\eZ_p  = \begin{pmatrix}0& 0& -\bar{p}\\ 0& 0& 0\\ p& 0 &0\\\end{pmatrix} ,
\nonumber\\
\eS_q &= \begin{pmatrix}q& 0& 0\\ 0& q& 0\\ 0& 0& -2q\\\end{pmatrix} ,
\label{XYZ}
\end{align}
where $p\in\OO'\otimes\OO$ and $q\in\Im\OO+\Im\OO'$.  It is straightforward to
work out the commutators of these elements of $\ee_8$ using matrix
multiplication \textit{except} for commutators involving two imaginary labels,
both from $\OO$, or both from~$\OO'$.  As described in~\cite{Lorentz}, and
discussed further in~\cite{Magic}, composition of imaginary elements of $\OO$
requires \textit{nesting} due to the lack of associativity.  We must therefore
introduce the additional elements
\begin{equation}
\eD_{p,q} = \frac12 \> [\eD_p,\eD_q] = \frac12 \> [\eX_p,\eX_q] ,  
\label{DD}
\end{equation}
where $p,q\in\Im\OO$ or $p,q\in\Im\OO'$.  We will normally assume that $p$,
$q$ are orthogonal and have unit norm.
There are $21+21=42$ such elements, generating $\so(7)\oplus\so(3,4)$, but only the
$14+14=28$ elements of $\gg_2+\gg_{2(2)}$ are independent of the matrices
given in~(\ref{XYZ}).  Specifically, due to triality, the $\eS_q$ can be
expressed in terms of the $\eD_{p,q}$.  We choose a particular basis
$\{\eG_q,\eA_q\}$ for the remaining nested elements, with
$\{\eA_i,...,\eA_\ell,\eG_k\}$ generating the $\su(3)\subset\gg_2$ that fixes
$k$.  These elements can be identified with the usual Gell-Mann matrices, with
$\{\eA_k,\eG_k\}$ corresponding (up to scale and a factor of $i$) to
$\{\lambda_3,\lambda_8\}$, respectively.  A similar construction holds in
$\OO'$, where however we fix $L$.  As shown in~\cite{Magic}, the resulting
algebra closes; the 248 elements given above form a basis for $\ee_8$ (using
\textit{either} the double-index $\eD$s, \textit{or} the $\eS$s, $\eG$s, and
$\eA$s).

As suggested by the form of the $\eD$s, we treat the upper left $2\times2$
block as special; the $64+14+42=120$ $\eX$s, $\eD$s, and double-index $\eD$s
generate $\so(12,4)$.  The remaining $64+64=128$ $\eY$s and $\eZ$s form a
single Mayorana--Weyl representation of $\so(12,4)$ under the adjoint action
in $\ee_8$, and will be called \textit{spinors} of $\ee_8$.  That is, we have
the decomposition
\begin{equation}
\ee_8 = \so(12,4) \oplus \mb{128} .
\end{equation}

The commutator action of $\so(12,4)$ on spinors, regarding both as elements of
$\ee_8$, is equivalent to (possibly nested) matrix multiplication, with the
$2\times2$ blocks of the $\so(12,4)$ matrices above acting on 2-component
spinors of the form
\begin{equation}
\alpha_\psi = \begin{pmatrix}-\bar{q}\\ p\\ \end{pmatrix}
\label{spinors}
\end{equation}
corresponding to the $\ee_8$ element $\psi = Y_p+Z_q$, with
$p,q\in\OO'\otimes\OO$.

A similar choice of preferred $2\times2$ block inside $3\times3$ matrices is a
feature of our earlier work~\cite{Schray,York,Denver} in a somewhat different
context, based on ideas suggested to us by Fairlie and Corrigan~\cite{FC}.
The essential idea is that vectors are squares of spinors
($X=\psi\psi^\dagger$), so the actions of spin groups on spinors
($\psi\longmapsto M\psi$) and on vectors ($X\longmapsto MXM^\dagger$) use the
same matrices.  Embedding these $2\times2$ matrices as above, and
reinterpreting both vectors ($2\times2$) and spinors ($2\times1$) as pieces of
$3\times3$ matrices results in a single $3\times3$ action that correctly
reproduces both the spinor and vector actions.

\section{The Enveloping Algebra}
\label{envelop}

We work throughout in the \textit{enveloping algebra} of $\ee_8$, that is,
the algebra of compositions of Lie algebra elements.  Such compositions must
always act on some representation of the given Lie algebra.  In many cases, we
restrict the adjoint action of $\ee_8$ (on itself) to that of $\so(12,4)$ on
the spinor $\mb{128}$.  Thus, for any elements $P,Q\in\so(12,4)$ and spinor
$\psi$, we define
$P\circ Q$ by
\begin{equation}
(P\circ Q)[\psi] = \bigl[P,[Q,\psi]\bigr] .
\end{equation}
The Jacobi identity now ensures that
\begin{equation}
(P\circ Q-Q\circ P)[\psi] = \bigl[[P,Q],\psi\bigr] ,
\end{equation}
so that $P\circ Q=Q\circ P$ if $[P,Q]=0$.

We emphasize that composition ($\circ$) provides a product on the enveloping
algebra; we can multiply elements, not merely take their commutators.  This
product is associative, since nested operations are always evaluated from the
inside out.  So we can also construct anticommutators in the enveloping
algebra.  Thus, given any \textit{proper} orthogonal subalgebra
$\so(p,q)\subset\so(12,4)$, we can, first of all, find a vector representation
$V$ of $\so(p,q)$ in its orthogonal complement in $\so(12,4)$, and then
construct the full (real) Clifford algebra $\Cl(p,q)$ as $\Cl(V)$ acting on
the $\mb{128}$.  The Clifford identity $\{\gamma_m,\gamma_n\}=2g_{mn}$ then
takes the form $Q\circ Q=-|Q|^2$, that is,
\begin{equation}
\bigl[Q,[Q,\psi]\bigr] = -|Q|^2\psi
\label{Clifford}
\end{equation}
for any $Q\in V$ and spinor $\psi$, with $|Q|^2$ denoting the Lorentz norm.
(The minus sign comes from our chosen Lorentz signature.)

The enveloping algebra can be used to construct the operation $\vp$, first
introduced in Section~\ref{so3}, which maps $L$ to $-L$ and also $IL$ to
$-IL$, etc.  This operation $\vp$ of ``$L$ conjugation'' can be realized on
$\Sp_\pm$ in the enveloping algebra of $\so(4,4)\subset\ee_8$ as
\begin{equation}
\vp = \eD_{KL}\circ \eD_{JL}\circ \eD_{IL}\circ \eD_L .
\end{equation}
However, since $\vp$ maps between different representations of $\so(9,1)$, it
can not be represented within (the enveloping algebra of) $\so(9,1)$ itself.

\section{The Killing Form}
\label{kill}

Every semisimple Lie algebra possesses a nondegenerate, symmetric inner
product, the \textit{Killing form} $B$.  For matrix Lie algebras, $B(M,N)$ can
be taken to be $\tr(MN)$.  The Killing form is unique up to an overall scale,
which we choose so that normalized boosts and rotations in $\so(12,4)$ square
to $\pm1$, respectively, when acting on spinors.  The basis for $\ee_8$ given
by omitting the $\eS_q$ from~(\ref{XYZ}) and~(\ref{DD}) is orthonormal (so
long as $p$, $q$ are normalized), as is the alternative basis using
$\{\eS_q,\eA_q,\eG_q\}$ instead of~(\ref{DD}) apart from the conventional
normalizations $B(\eS_q,\eS_q)=\pm6$, $B(\eG_q,\eG_q)=\pm3$,
$B(\eA_q,\eA_q)=\pm2$.
The \textit{signature} of (a real form of) a Lie
algebra is the number of boosts minus the number of rotations in any
orthonormal basis.  For example, the 26 boosts in (the half-split form of)
$\ee_6\subset\ee_8$ are the elements containing $L$ in their label
($\eX_{aL}$; $\eY_{aL}$; $\eZ_{aL}$; $\eD_L$; $\eS_L$), so the signature is
$26-52=-26$.  In $\ee_8$, the boosts are the $4\times8\times3$ $\eX$s, $\eY$s,
and $\eZ$s labeled by $IL$, $JL$, $KL$, $L$, as well as the $4\times4$ boosts
in the $\so(4,4)$ that act on labels in $\OO'$; linear combinations of the
latter include $\eD_L$ and $\eS_L$.  Thus, there are
$4\times8\times3+4\times4=112$ boosts in $\ee_8$, so the signature is


Given a nondegenerate, symmetric inner product $g$ and a complex structure
$\iota$ which is \textit{compatible} with $g$ in the sense that
\begin{equation}
g\bigl(\iota(\alpha),\iota(\beta)\bigr) = g(\alpha,\beta),
\label{compat}
\end{equation}
the inner product extends naturally to a hermitian product, of which it is the
real part.  Explicitly, the associated hermitian product is given by
\begin{equation}
\langle\alpha,\beta\rangle
  = g(\alpha,\beta) - i g\bigl(\alpha,\iota(\beta)\bigr) \in\CC
\label{hermit}
\end{equation}
with hermiticity following from~(\ref{compat}), which is equivalent to
\begin{equation}
g\bigl(\alpha,\iota(\beta)\bigr) = -g\bigl(\iota(\alpha),\beta\bigr) .
\end{equation}
(The complex unit $i$ in~(\ref{hermit}) is unrelated to the division algebra
structure used elsewhere in this paper.)  We could of course set $g=B$, but
this construction also applies to the inner product
\begin{equation}
g(\alpha,\beta) = B\bigl(\vp(\alpha),\beta\bigr)
\end{equation}
where $\vp$ denotes $L$ conjugation, as introduced in Section~\ref{envelop}.
The compatibility condition is still satisfied so long as $\vp$ commutes with
$\iota$, which it does for the complex structure $\iota=\eX_1$ introduced in
Section~\ref{so3}, when acting on spinors.  In this case, the pairing given by
$\vp$ is analogous to the dagger operation in traditional language, with
$\psi^\dagger\psi=\langle\psi,\psi\rangle=B(\vp(\psi),\psi)$.

\section{Volume Elements and Spinors}
\label{volume}

In this section, we use volume elements to describe how to decompose spinors
of $\ee_8$ into spinor representations of particular subalgebras of
$\so(12,4)$.

Every orthogonal Lie algebra $\so(m,n)$ is associated with an (abstract)
Clifford algebra $\Cl(m,n)$, whose degree-$1$ elements are the physicist's
gamma matrices, and whose degree-$2$ elements can be identified with
$\so(m,n)$ itself.  When $m+n$ is even, the Clifford product of any basis of
Cartan elements in $\so(m,n)$ is, up to an overall scale, independent of the
elements used to construct it; the resulting Clifford algebra element can be
thought of as the product of all of the gamma matrices.  We refer to this
normalized product as the \textit{volume element} $\omega_{m,n}$ of
$\Cl(m,n)$.  The volume element is a \textit{Casimir operator}, as it commutes
with all of $\so(m,n)$.

For the Lorentz Lie algebra $\so(3,1)$ and the weak Lie algebra $\so(4)$, the
volume elements $\omega_{3,1}$ and  $\omega_{4,0}$ square to $\mp1$ on spinors,
respectively.  In order to obtain real eigenvalues, we therefore define
\begin{equation}
\omega=\iota\circ\omega_{3,1},
\quad
\Omega=\omega_{4,0};
\end{equation}
$\omega$ can be thought of as $\gamma_5=i\gamma_0\gamma_1\gamma_2\gamma_3$.
The eigenvalues of the pair ($\omega$,$\Omega$) now provide binary labels for
the spinor $\mb{8}$s of $\so(3,1)\oplus\so(4)$.  These irreducible $\mb{8}$s
are the ``pots'' of spinors introduced in Section~\ref{construct}.
Furthermore, due to the presence of $\iota$ in $\omega$, we have
\begin{equation}
\omega_{9,1}=\omega\circ\Omega
\label{vol91}
\end{equation}
so that spinor representations of $\so(9,1)$ (such as $\Sp_\pm$ in
Section~\ref{e6}) can be distinguished by the sign of $\omega\circ\Omega$.

We emphasize that these representations depend only on the decomposition of
$\ee_6$ described in Section~\ref{e6}, independent of any choice of a maximal
set of commuting elements, that is, independent of the choice of Cartan
subalgebras.

This construction extends to $\ee_8$, where each representation of
$\so(3,1)\oplus\so(4)$ will also carry color labels, according to their
eigenvalues under $\eG_L$ and $\eA_L$, the chosen Cartan elements in $\suc$.
We interpret spinor representations for which both color eigenvalues are~$0$
as leptons; these representations can be distinguished using their eigenvalues
under either $\omega_{9,1}$ or $\eS_L$.  An alternate labeling is discussed in
Appendix~\ref{binary}.

Equivalently, the volume element $\omega\circ\Omega$ of $\so(9,1)$ divides the
$\mb{128}$ of $\so(12,4)$ into two $\mb{64}$s, which we denote $\Sp_\pm$ (thus
further generalizing the notation in Section~\ref{e6}).  Since there is only
one Majorana--Weyl representation of $\so(12,4)$ in $\ee_8$, the $\so(12,4)$
volume element $\Omega_{12,4}$ must act as a constant on this representation,
and hence on all spinors.  Similarly to~(\ref{vol91}), we also have
\begin{equation}
\Omega_{12,4} = \Omega_{9,1}\circ\Omega_{3,3}
\end{equation}
so that the actions of $\Omega_{9,1}$ and $\Omega_{3,3}$ on spinors are the
same.  Thus, $\Sp_\pm$ are also representations of $\so(3,3)$; they are, in
fact, irreducible representations of $\so(9,1)\oplus\so(3,3)$.

A further application of these volume elements arises when removing a complex
structure ($\iota\in\so(2)$) from $\so(m,n)$ yielding both $\so(m-2,n)$ and
\textit{two} vector representations of $\so(m-2,n)$, related by $\iota$.  On
any representation of $\so(m,n)$, the volume element
$\omega_{m,n}=\omega_{m-2,n}\,\omega_{2,0}$ must be constant, so that
$\omega_{m-2,n}=\pm\omega_{2,0}$ is just the complex structure itself.  In
other words, either one of the two vector representations can be chosen as the
degree-$1$ elements that generate $\Cl(m-2,n)$, with the other representation
then being the pseudovectors, that is, the elements of degree $m+n-3$.

\section{A Spinor Binary Code}
\label{binary}

We now choose specific Cartan elements in $\so(3,1)$ and $\so(4)$, then use
them to provide a binary code for the spinors, analogous to that of
Zee~\cite{Zee}.

For $\so(3,1)$, we associate the label $k$ with the $z$ direction.  The
resulting Cartan elements are the $xy$-rotation ($\eD_{i,j}$) and the
$zt$-boost ($B_z=X_{kL}$).  When acting on spinors, we replace $\eD_{i,j}$
with $L_z=\iota\circ D_{i,j}$ to ensure real eigenvalues.  We interpret the
sign of the eigenvalue of $L_z$ as giving the spin in the $z$ direction, as
usual.  The choice of $z$-direction to determine spin is, of course,
arbitrary.  We also have
\begin{equation}
\omega = L_z \circ B_z .
\end{equation}

We repeat this construction for $\so(4)$, with a few small but important
differences.  We begin by decomposing $\so(4)$ as $\su(2)\oplus\su(2)$.
Choosing $\so(3)\subset\so(4)$ to act only on the labels
$\{i\ell,j\ell,k\ell\}$, and pairing each resulting rotation
(e.g.\ $\eD_{i\ell,j\ell}$) with the unique generator of $\so(4)$ with which
it commutes, (e.g.\ $\eD_{k\ell,\ell}$), the sum and difference of these two
sets each generate an $\su(2)$.  As noted in Section~\ref{generations}, an
underappreciated
feature of this construction is that it does \textit{not} depend on the choice
of $\so(3)$; the two $\su(2)$ subalgebras are in fact unique, and are
precisely the subalgebras $\suR$ and $\suL$ introduced in
Sections~\ref{sl3H} and~\ref{e6}.  We interpret $\suL$ as the the
electroweak symmetry group, and, in the context of GUTs, $\suR$ as its
right-handed counterpart.

We now finally choose Cartan elements in $\suL$ and $\suR$, or,
equivalently, their sum and difference in $\so(4)$.  Since $\so(4)$ is
compact, every Cartan element will be a rotation, so we compose with $\iota$
when acting on spinors.  Our choice for the resulting Cartan elements of
(complexified) $\so(4)$ is
\begin{equation}
\eW_a = \iota\circ D_{i\ell,j\ell},
\quad
\eW_b = \iota\circ D_{k\ell,\ell},
\end{equation}
and our conventions are such that $\eW_z=\eW_a-\eW_b\in\suL\subset\gg_2$.
We can not overemphasize that this association of $\ell$ with $k\ell$ is
completely independent of the association of $L$ with $k$ in $\so(3,1)$,
although this combination does simplify the notation later.  We also reiterate
that the decomposition $\so(4)=\suL\oplus\suR$ is unique; it is unaffected by
the choice of Cartan elements in $\so(4)$.  In any case, we have
\begin{equation}
\Omega = - \eW_a \circ \eW_b = D_{i\ell,j\ell} \circ D_{k\ell,\ell} .
\end{equation}

\begin{table}
\centering
\begin{tabular}{|c|c|c|}
\hline
\textbf{Code} & \boldmath $\Re(\bpsi)$, $\Im(\bpsi)$ & \textbf{Name} \\
\hline
&&\\[-11pt]
\hline
$|+-++\rangle$ & $\eY_{1(\UU+L)}+\eZ_{k(\UU-L)}$, & $\nu_{\Ru}$ \\
& \quad $-\eY_{k(\UU+L)}-\eZ_{1(\UU-L)}$ & \\
\hline
$|+---\rangle$ & $\eY_{j(\UU+L)}+\eZ_{i(\UU-L)}$, & $\nu_{\Rd}$ \\
& \quad $-\eY_{i(\UU+L)}+\eZ_{j(\UU-L)}$ & \\
\hline
$|-+++\rangle$ & $-\eY_{j(\UU+L)}+\eZ_{i(\UU-L)}$, & $e_{\Ru}$ \\
& \quad $-\eY_{i(\UU+L)}-\eZ_{j(\UU-L)}$ & \\
\hline
$|-+--\rangle$ & $\eY_{1(\UU+L)}-\eZ_{k(\UU-L)}$, & $e_{\Rd}$ \\
& \quad $\eY_{k(\UU+L)}-\eZ_{1(\UU-L)}$ & \\
\hline
&&\\[-10pt]
\hline
$|+++-\rangle$ & $\eY_{k\ell(\UU+L)}+\eZ_{\ell(\UU-L)}$, & $\nu_{\Lu}$ \\
& \quad $-\eY_{\ell(\UU+L)}+\eZ_{k\ell(\UU-L)}$ & \\
\hline
$|++-+\rangle$ & $\eY_{i\ell(\UU+L)}-\eZ_{j\ell)(\UU-L)}$, & $\nu_{\Ld}$ \\
& \quad $\eY_{j\ell)(\UU+L)}+\eZ_{i\ell(\UU-L)}$ & \\
\hline
$|--+-\rangle$ & $\eY_{i\ell(\UU+L)}+\eZ_{j\ell(\UU-L)}$, & $e_{\Lu}$ \\
& \quad $-\eY_{j\ell(\UU+L)}+\eZ_{i\ell(\UU-L)}$ & \\
\hline
$|---+\rangle$ & $-\eY_{k\ell(\UU+L)}+\eZ_{\ell(\UU-L)}$, & $e_{\Ld}$ \\
& \quad $-\eY_{\ell(\UU+L)}-\eZ_{k\ell(\UU-L)}$ & \\
\hline
\end{tabular}
\caption{The binary code for the spinors in the $\mb{16}$ of $\so(9,1)$ with
$\omega\circ\Omega=1$, corresponding to two complex $\mb{4}$s of
$\so(3,1)\oplus\so(4)$ (above and below the double line), along with their
interpretation.
}
\label{code}
\end{table}

Each pot now admits a basis of simultaneous eigenstates of the Cartan elements
$\{\eW_a,\eW_b,L_z,B_z\}$, with eigenvalues $\pm1$ in each case.  We therefore
have a four-digit binary code $b_1b_2b_3b_4$ labeling the 16 independent
(complex!)\ spinor degrees of freedom, with $b_m=\pm1$.  The product of all
four digits is constant within a ``double pot'' $\Sp_\pm$, and either the
product of the first two or last two digits can be used to distinguish the two
pots within a given double pot.

A table listing the binary codes of the $\ee_6$ spinors in $\Sp_+$ (so
$\omega\circ\Omega=1$) is given in Table~\ref{code}, along with their
interpretation.  For each pair of basis elements given in the table,
$\iota\bigl(\Re(\psi)\bigr)=\Im(\psi)$, so that the complex spinor
$(a+bi)\psi$ becomes $a\,\Re(\psi)+b\,\Im(\psi)$ when represented within
$\ee_6$.

This binary code is similar in spirit to the five-digit code of
Zee~\cite{Zee}.  The spinors shown in Table~\ref{code} are all leptons, so no
color labels are needed.  As discussed in Appendix~\ref{volume}, color is
determined by the eigenvalues of $\eG_L$ and $\eA_L$, although $\eS_L$ is also
needed to distinguish leptons from antileptons.  Using the equivalent Cartan
basis $\{\eD_{I,IL},\eD_{J,JL},\eD_{K,KL}\}$, 
each with eigenvalues $\pm1$,
reproduces the first three digits of the Zee binary code.  Our resulting
seven-digit code then matches that of Zee, with an additional two digits at
the end associated with the two Lorentz Cartan elements.

This seven-digit code labels 128 possible states, yet there are only 64
(complex!)\ spinor states available to us.  How is this possible?  The
seven-digit code corresponds to seven of the eight Cartan elements of
$\so(12,4)$.  However, three of the seven were originally rotations, requiring
composition with $X_1$, the eighth Cartan element.  Thus, the product of the
digits of our seven-digit code yields (minus) the eigenvalue under the product
of \textit{all eight} Cartan elements, that is, of the volume element
$\Omega_{12,4}$.  Since we have a single, Majorana--Weyl representation of
$\so(12,4)$, this eigenvalue must be constant; with our conventions, it is
$+1$.  Thus, the seventh digit can in fact be dropped, and we recover
precisely the Zee five-digit code with one additional digit for spin.

\onecolumngrid
\clearpage
\goodbreak
\twocolumngrid
\bibliography{octo}

\end{document}